\documentclass[aps,preprint,amsmath,amssymb,superscriptaddress,floatfix]{revtex4}
\usepackage{amsfonts}
\usepackage{amsmath}
\usepackage{amssymb}
\usepackage{graphicx}%
\usepackage{ulem}
\usepackage{color}
\usepackage{epstopdf}
\usepackage{microtype}

\providecommand{\U}[1]{\protect\rule{.1in}{.1in}}

\begin{document}

\title{Non-equilibrium  quantum phase transition via entanglement decoherence dynamics}

\author{Yu-Chen Lin}
\affiliation{Department of Physics and Centre for Quantum Information Science, National Cheng Kung University, Tainan 70101, Taiwan}
\author{Pei-Yun Yang}
\affiliation{Department of Physics and Centre for Quantum Information Science, National Cheng Kung University, Tainan 70101, Taiwan}
\author{Wei-Min Zhang}
\email{wzhang@mail.ncku.edu.tw}
\affiliation{Department of Physics and Centre for Quantum Information Science, National Cheng Kung University, Tainan 70101, Taiwan}

\begin{abstract}
\textbf{We investigate the decoherence dynamics of continuous
variable entanglement as the system-environment coupling strength
varies from the weak-coupling to the strong-coupling regimes.  Due
to the existence of localized modes in the strong-coupling regime,
the system cannot approach equilibrium with its environment, which 
induces a nonequilibrium quantum phase transition.  We analytically
solve the entanglement decoherence dynamics for an arbitrary
spectral density.  The nonequilibrium quantum phase transition is
demonstrated as the system-environment coupling strength varies for
all the Ohmic-type spectral densities. The 3-D entanglement quantum
phase diagram is obtained.}
\end{abstract}


\maketitle

In quantum many-body systems, quantum phase transitions (QPTs) may occur at zero temperature
as a parameter varies in the Hamiltonian of the system, induced purely by quantum fluctuations \cite{Sachdev2011}.
QPTs have been explored via entanglement \cite{Osterloh2002,Osborne2002,Vidal2003,Gu2004,Wu2004},
because entanglement is regarded as a key resource to detect QPTs, owing to the fact that
the entanglement can exist without any classical correlations \cite{Kaszlikowski2008}. In
previous investigations, QPTs are usually investigated in terms of entanglement for the many-body
systems via the von Neumann entropy by dividing the system into various bipartites. In these investigations,
significant changes of the entanglement as a parameter varies in the Hamiltonian of the system provide
a deeper understanding about QPTs. 

In order to experimentally explore QPTs, it is crucial how to manipulate the basic parameters in the Hamiltonian
of the system, such as hopping energies and interaction coupling strengthes, etc. through the external devices.
Thus, these systems manifesting QPTs become naturally open systems \cite{Verstraete2009, Mitra2006, Diehl2008}.
In this article, we attempt to explore nonequilibrium QPTs by studying the entanglement
decoherence dynamics in a prototype open quantum system of two entangled modes interacting with a general 
non-Markovian environment.
We find that entanglement decoherence dynamics, induced by the interaction between the system
and its environment, manifests a significant change as the system-environment coupling strength
varies from the weak to the strong coupling regimes. Nonequilibrium quantum phase transition
occurs due to the competition between quantum dissipation dynamics and localization. 
Thermal fluctuations can drive further the entanglement decoherence dynamics into a quantum critical 
regime and then into the thermal disordered regime. This could open a new venue for the experimental 
investigations of QPTs through the real-time entanglement decoherence dynamics.

\vspace{0.5cm}
\noindent \textbf{\large Results}

\noindent \textbf{Real-time exact solution of entangled
squeezed states}. We consider a system with two entangled
continuous variables, such as two entangled cavity fields,
interacting to a common thermal environment, its dynamics is
described by the Fano-Anderson Hamiltonian
\cite{Anderson1961,Fano1961},
\begin{align}
H_{tot} 
= &~ \omega_{1}a_{1}^{\dagger}a_{1}+\omega_{2}a_{2}^{\dagger}a_{2}
+ \kappa(a_{1}^{\dagger}a_{2}+a_{2}^{\dagger}a_{1}) \notag \\
& + \sum_{k} \omega_{k}b_{k}^{\dagger}b_{k}
+ \!\!\! \sum_{i=1.2,k}\!\!(g_{ik}a_{i}^{\dagger}b_{k}+g_{ik}^{\ast}b_{k}^{\dagger}a_{i}),
\end{align}
where $a_{i}$ and $a_{i}^{\dagger}$(i=1,2) are the annihilation
and creation operators of the two continuous variables with frequency $\omega_{i}$, and $\kappa$
is a real coupling constant between the two continuous variables, which can be tuned, for example,
through a beam splitter on the two single-mode fields \cite{Cassettari2000}. The environment
Hamiltonian consists of an infinite number of bosonic modes, where $b_{k}$ and $b_{k}^{\dagger}$
are the annihilation and creation operators of the mode $k$ with frequency $\omega_{k}$.
The interaction between the system and the environment is given by the last term in Eq.~(1),
and the parameter $g_{ik}$ is the coupling amplitude between the continuous variable mode $i$
and the environment mode $k$.  The complexity of the problem is embedded in the spectral density,
see Eq.~(\ref{osp}) later, which characterizes the complicated energy structure of the environment plus the
system-environment interaction.

In order to investigate the entanglement decoherence dynamics of the two continuous
variables under the influence of a complicated environment, we take a decoupled initial state
between the system and the environment \cite{Leggett1987}: $\rho_{tot}(0)\! =\!
\rho_s(0) \otimes \rho_E(0)$. The initial state of the system is $\rho_s(0) \!=\! |\psi_s(0)\rangle
\langle \psi_s(0)|$, where $|\psi_s(0)\rangle$ is an entangled state between the two
continuous variables, and the
environment is initially in thermal equilibrium $\rho_E(0) \!=\! \frac{1}{Z}\exp\{-\beta \!\sum_{k} 
\omega_{k}b_{k}^{\dagger}b_{k}\}$ at the initial inverse temperature $\beta=1/k_BT$.
After the initial time, the system and the environment both dynamically evolve into non-equilibrium states. 
To be specific, let $\rho_s(0)$ be a two-mode squeezed state \cite{Klauder86},
\begin{align}
\label{ins}
|\psi_s(0)\rangle=S_{12}(r)|00\rangle,
\end{align}
where $S_{12}(r)=\exp\{r
a_1a_2-ra_{1}^{\dagger}a_{2}^{\dagger} \}$ is a two-mode squeezed operator, and $r$ is the real
squeezing parameter.  This state has been experimentally realized in many different systems, first 
given by Heidmann \textit {et al} \cite{Heidmann1987}, and has also been applied to quantum 
teleportation \cite{Braunstein1998}. When the squeezing parameter becomes very
large, the above state would approach to the ideal
Einstein-Podolsky-Rosen (EPR) state \cite{Einstein1935}.

Without loss of generality, we may consider the two continuous variables as two identical fields
and interact homogeneously with the environment, namely, $\omega_{1}=\omega_{2}=\omega_{0}$
and $g_{ik}=g_{k}$.
Then we can introduce the center-of-mass and the relative motional variables, respectively,
given by $a_{+}^{\dagger}=(a_{1}^{\dag}+a_{2}^{\dagger})/\sqrt{2}$, and $a_{-}^{\dagger}
=(a_{1}^{\dag}-a_{2}^{\dagger})/\sqrt{2}$, with the corresponding frequencies $\omega_\pm
= \omega_0 \pm \kappa$. It is easy to check that only the center-of-mass
variable $a_{+}$ is coupled with the environment, and the relative-motional variable decouples
from the environment which forms a decoherence-free subspace \cite{Lidar1998}. If the two continuous
variables are not identically coupled to the environment, no such decoherence-free
subspace exists, and the decoherence dynamics of the relative motion  behaves similarly as that
of the center-of-mass motion.
In terms of the center-of-mass motion and the relative motion of the two continuous variables,
we can rewrite the initial state (\ref{ins}) as a direct product of two entangled states,
$|\psi_s(0)\rangle=|\psi_{+}(0)\rangle\otimes|\psi_{-}(0)\rangle$, where
\begin{align}
|\psi_{\pm}(0)\rangle=\exp\Big[\pm\frac{r}{2}(a_{\pm})^2\mp\frac{r}{2}(a^\dag_\pm)^2\Big]|0\rangle.
\end{align}
Because the relative motion $a_{-}$ is decoupled from the
environment, the entanglement between the two continuous variables
in $|\psi_{-}(0)\rangle$ is decoherence-free. Therefore, the
entanglement decoherence only happens to the
center-of-mass motion. 

More specifically, entanglement decoherence dynamics is fully determined by the spectral density of the environment,
which is defined by
$J_{ij}(\omega)=\sum_k g_{ik}g^*_{jk} \delta(\omega-\omega_k)$.  When the two continuous variables 
identically couple to the common environment: $g_{ik}=g_{k}$, we have $J_{ii}(\omega) =|J_{12}(\omega)|$. 
In the center-of-mass frame, the spectral density is simply reduced to
\begin{align}
\boldsymbol{J}(\omega)&=\left(
\begin{array}[c]{cc}
J_{++}(\omega) & J_{+-}(\omega) \\
J_{-+}(\omega) & J_{--}(\omega)
\end{array} \right)=\left(
\begin{array}[c]{cc}
2J(\omega) & 0 \\
0 & 0
\end{array} \right),
\end{align}
it shows that $J_{--}(\omega)=0$. This indicates that the environment synchronizes   
quantum as well as thermal fluctuations at the two identical parties, so that the relative 
motion experiences no fluctuation. Consequently, the corresponding entanglement contribution remains 
unchanged as the initial value for the relative motion (i.e. decoherence-free), as we will see more 
discussion later.  The decoherence dynamics of the center-of-motion part is then fully controlled by the spectral
component $J_{++}(\omega)=2J(\omega)$.
In the following, we consider the Ohmic-type spectral density~\cite{Leggett1987} which
can simulate a large class of thermal bath,
\begin{align}
\label{osp} 
J(\omega)&=\sum_k |g_{k}|^2 \delta(\omega-\omega_k)=
\eta\omega(\frac{\omega}{\omega_{c}})^{s-1}\exp(-\frac{\omega}{\omega_{c}}),
\end{align}
where $\eta$ is a dimensionless coupling strength between the system
and the environment, and $\omega_{c}$ is the cutoff of the
environment spectrum. The parameter $s$ classifies the environment
as sub-Ohmic $(s<1)$, Ohmic $(s=1)$, or super-Ohmic $(s>1)$. Although we focus on such a general environmental structure, the
results are valid to other spectral densities (other environmental
structures)
that do not be described by Eq.~(\ref{osp}).

The real-time dynamics of the initial state (\ref{ins}) can be solved 
directly from the reduced density matrix: $\rho_s(t) \!=\! {\rm Tr_E}[{\cal U}(t,0)\rho_{tot}(0){\cal U}^\dag(t,0)]$,
where ${\cal U}(t,0) \!=\! \exp\{-iH_{tot}t\}$ is the unitary evolution operator of the total system 
(the two entangled fields coupled their environment).
The general solution is given by
\begin{align}
\label{separate rho}
\rho_s(t)=\rho_+(t) \otimes  \rho_{-}(t)
\end{align}
where $\rho_{+}(t)$ is the reduced density matrix of the
center-of-mass motion, and the relative motion remains a pure
state $\rho_-(t)= |\psi_-(t)\rangle \langle \psi_-(t)|$.  
The explicit analytical solution can be solved with the results
\label{rhot}
\begin{align}
&|\psi_-(t)\rangle=S[r_-(t)]
|0\rangle, \\
&\rho_{+}(t)=S[r_+(t)]\rho_{\rm th}(t)S^\dag[r_+(t)]
\label{rho+t}
\end{align}
where $S[r_\pm(t)]=\exp\Big[ \frac{r^*_\pm(t)}{2}(a_\pm)^2 -
\frac{r_\pm(t)}{2}(a_\pm^\dag)^2\Big]$ is the time-evolving
squeezing operators of the center-of-mass motion and the relative
motion of the two fields, and Eq.~(\ref{rho+t}) is a time-evolving thermal-like
squeezed state \cite{Tan2011}.
Explicitly, the squeezing parameters 
\begin{align}
r_-(t) = -r e^{-i2 \omega_- t} ~,~~ r_+(t)=|r_+(t)|e^{i\theta_+(t)}
\end{align}
with 
$|r_+(t)|=\frac{1}{4}\ln\frac{n_+(t)+|\sigma_+(t)|+1/2}{n_+(t)-|\sigma_+(t)|+1/2}$
and $\theta_+(t)=\arg[\sigma_+(t)]$. 
The nonequilibrium thermal-like state
\begin{align}
\rho_{\rm th}(t)=\sum_n\frac{[\overline{n}_+(t)]^n}{[1+\overline{n}_+(t)]^{n+1}}
|n\rangle_+ {}_+\langle n|,  \label{sts}
\end{align}
where $|n\rangle_+ = \frac{1}{\sqrt{n!}} (a^\dag_+)^n |0\rangle$ is the Fock state of the 
center-of-mass variable, and the average particle number in this state, 
$\overline{n}_+(t)={\rm Tr}[a^\dag_+a_+ \rho_{\rm th}(t)]$, describes the nonequilibrium
thermal fluctuations and satisfies the relation
\begin{align}
\overline{n}_+(t) + 1/2= \sqrt{[n_+(t)+1/2]^2-|\sigma_+(t)|^2} .  \label{rst}
\end{align}
The factor $1/2$ in the above equation is related to the zero-point energy fluctuation,
and $n_+(t)$ and $\sigma_+(t)$ are respectively the photon intensity (the squeezed thermal fluctuation) 
and the two-photon correlation of the center-of-mass variable, 
\label{ns}
\begin{align}
n_+(t)=&{\rm Tr}[ a^\dag_+ a_+ \rho_+(t)] =|u^2_+(t,0)|\sinh^2\!r+v_+(t,t),  \label{n+}\\
\sigma_+(t)=&{\rm Tr}[ a_+a_+ \rho_+(t)]=-u^2_+(t,0)\cosh\!r \sinh\!r.   \label{s+}
\end{align}
The functions $u_{+}(t,0)$ and $v_{+}(t,t)$ are Schwinger-Keldysh's
retarded and correlated (fluctuation) Green functions in nonequilibrium many-body systems
\cite{Schwinger1961,Kadanoff1962,Keldysh1965} for the center-of-mass variable, and they obey the Kadanoff-Baym equations
\cite{Kadanoff1962} and the dissipation-fluctuation theorem \cite{Zhang2012},
respectively,
\begin{align}
&\dot{u}_{+}(t,0)+i\omega_+u_{+}(t,0)+ \!\! \int_0^t  \!\!\! d\tau g(t,\tau)u_{+}(\tau,0)=0,  \label{u+t} \\
&v_+(t,t)=\int_0^{t}\!\!d\tau_{1}\!\!\int_0^{t} \!\!d\tau_{2}u_+(t,\tau_{1})
\widetilde{g}(\tau_{1},\tau_{2})u_+^{\dagger}(t,\tau_{2}) , \label{v+t}
\end{align} 
in which the integral kernels, $g(t,t')\!=\!2\int d\omega J(\omega)
e^{-i\omega(t-t')}$ and $\widetilde{g}(t,t')=2\int d\omega J(\omega) \overline{n}(\omega, T)e^{-i\omega(t-t')}$,
are fully determined by the spectral density of the environment. Here, $\overline{n}(\omega, T)=1/[e^{\beta \omega}-1]$ 
is the initial particle distributions in the environment.

The above analytical solutions show explicitly that the squeezed parameter in the relative-motion state 
$\rho_-(t)$ only takes a simple oscillation and is therefore decoherence-free. This is because 
the environment which equally couples to the two identical fields synchronizes both quantum 
(dissipation) and thermal fluctuations at the two identical parties so that the relative motion 
experiences no fluctuation, as a consequence of  $J_{--}(\omega)=0$.   
On the other hand, all the environment-induced
quantum (dissipation) and thermal fluctuations derive the center-of-mass motion part from 
an initial pure squeezed state into a thermal-like squeezed state $\rho_+(t)$, see Eq.~(\ref{rho+t}) in which the
squeezed operator $S[r_+(t)]$ acts on the nonequilibrium thermal-like state $\rho_{th}(t)$.
The nonequilibrium thermal-like state $\rho_{th}(t)$, which is induced by the environment, is 
different from the usual equilibrium thermal state because the averaged particle number 
$\overline{n}_+(t)$ in $\rho_{th}(t)$ is not equal to the standard Bose-Einstein distribution 
$\overline{n}(\omega, T)$ at the given frequency $\omega$  and the given equilibrium 
temperature $T$. Geometrically, $\rho_{th}(t)$ is also symmetrically distributed in terms 
of the quadrature components of the center-of-mass variable with  $\Delta X_+(t)
=\Delta P_+(t)=\sqrt{\overline{n}_+(t)+1/2}$, where $\overline{n}_+(t)$ and $1/2$ characterize 
the thermal and vacuum fluctuations, respectively. The squeezed parameter $r_+(t)$ is 
governed by both the quantum fluctuation $\sigma_+(t)$ and the thermal fluctuations 
embedded in the intensity $n_+(t)$ of the center-of-mass variable. The relation given by 
Eq.~(\ref{rst}), $n_+(t)+1/2 = \sqrt{[\overline{n}_+(t) +1/2]^2 + |\sigma_+(t)|^2}$ describes
how the thermal-like state $\rho_{th}(t)$, is squeezed by the quantum fluctuation 
$\sigma_+(t)$ through the squeezing operator $S(r_+(t))$:  $\Delta X_+(t) \rightarrow 
\sqrt{\overline{n}_+(t)+1/2}\,e^{-|r_+(t)|}$ and $\Delta P_+(t) \rightarrow \sqrt{\overline{n}_+(t)
+1/2}\, e^{|r_+(t)|}$.  If $\sigma_+(t) \rightarrow 0$,  we have $r_+(t) \rightarrow 0$ and $S(r_+(t)) 
\rightarrow 1$. Then the center-of-mass motion approaches to a thermal state due to 
thermal fluctuations of the environment only.

\vspace{0.5cm}

\noindent \textbf{Nonequilibrium entanglement decoherence dynamics}.

The above analytical solutions show that the entanglement decoherence dynamics
can be fully determined from the solution of Eq.~(\ref{u+t}) which provides the general quantum 
dissipation dynamics in open quantum systems \cite{Zhang2012},
\begin{align}
\label{u++t} u_{+}(t,0)=& \sum_jZ_je^{-i\overline{\omega}_j t} \!+2\!\!\int
\!\!\frac{d\omega J(\omega)e^{-i\omega t}}{(\omega \!-\!\omega_+
\!-\!\!\Delta(\omega))^{2}\!+\!4\pi^2J^{2}(\omega)} ,
\end{align}
where the first term is contributed by the localized modes in the
Fano-Anderson model \cite{Mahan2000,Zhang2012}, with the frequencies
$\overline{\omega}_j$ which are determined by the zeros of the function $z(\omega) \equiv 
\omega_+ -\omega +\Delta(\omega)$, and the amplitude
$Z_j=1/(1-\Sigma'(\overline{\omega}_j))$ is the corresponding pole residue. 
Here $\Sigma(\omega)=2\! \int
\!d\omega' \frac{J(\omega')}{\omega-\omega'}$ is the self-energy,
and $\Delta(\omega)=2{\cal P}\!\!\int \!
d\omega'\frac{J(\omega')}{\omega-\omega'}$ denotes its principal
value. This localized mode contributes a dissipationless dynamics. The second term
in Eq.~(\ref{u++t}) always decays, which leads to the quantum
dissipation (damping) dynamics in open systems. For the spectral
density of Eq.~(\ref{osp}), the localized mode occurs only when the
dimensionless system-environment coupling strength is greater than the 
critical coupling strength $ \eta_c(s)$ for a given environment 
characterized by $s$, 
\begin{align}
\eta > \eta_c(s) \equiv \frac{\omega_+}{2\omega_{c}\Gamma(s)} , \label{crip}
\end{align}
where $\Gamma(s)=\int_0^\infty x^{s-1}e^{-x}dx$ is the gamma function.
On the other hand, the environment-induced fluctuations  is characterized by $v_+(t,t)$,
which is determined by Eq.~(\ref{v+t}) as a result of
the generalized nonequilibrium fluctuation-dissipation theorem in the time-domain.

Now we can see that the center-of-mass state $\rho_{+}(t)$ contains various
decoherence dynamics.  For a given spectral density  (fixed $s$),
the state $\rho_{+}(t)$ in the strong-coupling regime ($\eta \!>\!
\eta_c(s)$) undergoes a partial decoherence process and then
approach to a nonequilibrium state, due to the existence of the localized mode
\cite{Mahan2000,Anderson1958,Cai2014,Xiong2015,Lo2015,Nan2015}. This
property becomes obvious by looking at the initial state dependence,
the squeezed parameter $r$-dependence in the time-dependent
coefficients in Eqs.~(\ref{n+}-\ref{s+}).  This initial-state dependence is a
manifestation of the strong non-Markovian memory effect, induced by
the localized mode in the strong coupling, see Eq.~(\ref{u++t}).
Thus, when $\eta \!>\! \eta_c(s)$, $\rho_{+}(t)$ becomes a
nonequilibrium entanglement state that always depends on the initial
state, even in the steady-state limit $t\!\rightarrow\!\infty$. Only
in the weak-coupling regime ($\eta \!<\!\eta_c(s)$), the localized
mode cannot be formed, and the Green function $u_{+}(t,\!0)
\!\rightarrow \!0$ as $t\!\rightarrow \!\infty$ so that $\sigma_+(t)
\!\rightarrow \!0$. Then the time-dependent squeezing parameter
$r_+(t)\rightarrow 0$. As a result, the state $\rho_{+}(t)$ will
approach to thermal equilibrium state,
\begin{align}
\label{thermal state}
\rho_{+}(t\rightarrow\infty) = \rho_{\rm th}(t\rightarrow \infty) 
\end{align}
with the average particle number (thermal fluctuation) $\overline{n}_+=n_+=v_+(t \rightarrow \infty)$.
This state is independent of the initial squeezed state,
and also contains no any entanglement. In other words, when
$\eta \!<\!\eta_c(s)$, the system reaches equilibrium with its environment, and
the entanglement of the center-of-mass state $\rho_{+}(t)$ is completely decohered
away.

To be explicit, we should quantitatively study the entanglement decoherence dynamics by quantifying
the entanglement between the two continuous variables in the two-mode entangled state, 
using the logarithmic negativity \cite{Vidal2002}.
The logarithmic negativity has been widely used in the literature, based on the Peres-Horodecki positive partial
transpose (PPT) criterion \cite{Simon2000} as a necessary and sufficient condition for the separability
of bipartite Gaussian states.  The
entanglement degree of a bipartite Gaussian state is given by \cite{Vidal2002}:
\begin{align}
E_{N}=max\{0,-log_{2}(2\tilde{\lambda}_{2})\}.\label{logarithmic negativity}
\end{align}
where $\tilde{\lambda}_{2}$ is the smaller one of the two symplectic eigenvalues of the covariant matrix.
On the other hand, if the entangled state is a direct product of two entangled states, the total entanglement is 
also the sum of the logarithmic negativities from each state  \cite{Vidal2002}.  Meanwhile, we also find that
for the case of the direct product state, like Eq.~(\ref{separate rho}), in which one state undergoes a 
decoherence process and the other state is decoherence-free, one must calculate the entanglement from each state 
separately in the product and then add them together to get the correct total entanglement. If one uses
the logarithmic negativity to calculate the entanglement directly from the original state, then the decoherence
dynamics in one state will artificially decohere away the entanglement in the decoherence-free state. This will
lead to a unphysical solution.
Thus, the total entanglement between the two original continuous variables in $\rho_s(t)$ is given by
\begin{align}
E_{N}=E_{N}(\rho_{+}(t))+E_{N}(\rho_{-}(t)).
\label{additional property}
\end{align}
It is not difficult to find that for the initial state (\ref{ins}), 
the entanglement $E_{N}(\rho_{+}(0))=E_{N}(\rho_{-}(0))=r/ln2$,
namely, the initial total entanglement is equally distributed between the relative-motion
state $|\psi_{-}(0)\rangle$ and the center-of-mass state $|\psi_{+}(0)\rangle$.

The real-time dynamics of the entanglement for each part in Eq.~(\ref{additional property}) can be solved analytically.
For the relative-motion state $\rho_{-}(t)$, because of its decoupling from the thermal
reservoir, its entanglement remains {\it unchanged}, $E_{N}(\rho_{-}(t))=r/ln2$.
On the other hand, for the center-of-mass motion state $\rho_{+}(t)$, the time evolution of the
entanglement  can be determined through Eq.~(\ref{logarithmic negativity}).
It is not difficult to find the smaller symplectic eigenvalue $\tilde{\lambda}_{2}(t)$ of the covariant
matrix with respect to the two original continuous variables for $\rho_{+}(t)$,
\begin{align}
\tilde{\lambda}_{2}(t)=\frac{1}{\sqrt{2}}\sqrt{n_+(t)-|\sigma_+(t)|+1/2}.
\label{smaller sym}
\end{align}
Then $E_{N}(\rho_{+}(t))$ can be computed from the nonequilibrium
retarded and correlated Green functions $u_+(t,0)$ and $v_+(t,t)$
through Eqs.~(\ref{n+}) and (\ref{s+}). The retarded Green function
$u_+(t,0)$ describes the dissipation dynamics, which is independent
of the initial temperature of the environment, and manifests pure
quantum correlations between the system and the environment \cite{Zhang2012}. The
correlation Green function $v_+(t,t)$, the generalized nonequilibrium
fluctuation-dissipation theorem, depends explicitly on the initial
environment temperature, it describes the thermal fluctuations
and the thermal-fluctuation-induced quantum fluctuations. The 
dynamics of the total entanglement of the two
continuous variables with the initial state (\ref{ins}) is given by
$E_{N}=E_{N}(\rho_{+}(t))+r/ln2$.

We first consider the weak-coupling regime  $\eta <\eta_c(s)$ at
zero temperature, the  function $u_+(t,t_{0})$ will decay to zero
for $t\rightarrow\infty$. Meanwhile, the correlation function
$v_{+}(t,t)=0$ because $\overline{n}(\omega,T)=0$ at zero temperature.
Thus the time-dependent functions in Eqs.~(\ref{n+}-\ref{s+}) lead to
$\sigma_+(t)\!=\!0$ and $n_+(t)\!=\!0$ as $t\!\rightarrow\! \infty$. As a result,
Eq.~(\ref{smaller sym}) is reduce to
$\tilde{\lambda}_{2}(t\!\rightarrow\! \infty)=\frac{1}{2}$ so that
$E_{N}(\rho_{+}(t\!\rightarrow\!\infty))=0$. Then the steady-state total
entanglement $E_{N}=E_{N}(\rho_{-}(t\!\rightarrow\!\infty))=r/ln2$. This reproduces the
result at zero temperature in the weak-coupling regime we obtained
previously \cite{An2007}. If the environment is initially at a
finite temperature, because
$u_{+}(t\!\rightarrow\!\infty,0)\!\rightarrow \!0$ is always true
for $\eta <\eta_c(s)$, we have $n_+(t)\! \rightarrow \!v_+(t,t)$ and
$\sigma_+(t) \!\rightarrow \!0$ so that
\begin{align}
\label{thermal state En}
-log_{2}(2\tilde{\lambda}_{2})=-\frac{1}{2ln2}ln(1+2v_{+}(t,t)).
 \end{align}
Also because $v_{+}(t,t)>0$ for any finite temperature,
Eq.~(\ref{thermal state En}) gives always a negative value.
According to the criterion (\ref{logarithmic negativity}), again the
entanglement of $\rho_{+}(t\rightarrow \infty)$ must go to zero.
Indeed, Eq.~(\ref{thermal state En}) shows that thermal fluctuations
speed up the entanglement decoherence. The steady-state entanglement
$E_{N}(\rho_{+}(t\rightarrow\infty))$ remains zero in the weak-coupling 
regime ($\eta <\eta_c(s))$ for any initial environment
temperature, and the steady-state total entanglement always equals
to $E_{N}(\rho_{-}(t))=r/ln2$.

Actually, the entanglement dynamics
of Eq.~(\ref{ins}) in the weak-coupling regime at finite temperature
was previously studied in Ref.~\cite{Paz2009} where
it shows that the decoherence lets the total entanglement be less than
one half of the initial total entanglement due to the thermal effect, see 
explicitly Fig.~8(b) in Ref.~\cite{Paz2009}.
This is obviously a wrong result because the relative motion state
$\rho_{-}(t)$ decouples from the environment such that the entanglement
contained in $\rho_{-}(t)$ is decoherence-free.
Thus, physically the steady-state total entanglement can never be less than
$r/ln2$, as we shown above. The mistake made in Ref.~\cite{Paz2009}
comes from an improper calculation of the entanglement when the entangled 
state is a direct product of two states.  The improper calculation in Ref.~\cite{Paz2009} 
lets the entanglement decoherence dynamics of the center-of-mass motion  
artificially decohere the entanglement in the relative motion state, while the later is however decoherence-free. 
This leads to the wrong result of the steady-state total entanglement being less than
$r/ln2$, as given in Ref.~\cite{Paz2009}.  

On the other hand, in the strong-coupling regime ($\eta>\eta_c(s)$),
the propagating function $u_+(t,t_0)$ will not decay to zero in the
steady-state limit, due to the existence of localized states
\cite{Zhang2012,Mahan2000}. Thus the state $\rho_{+}(t)$
Eq.~(\ref{rho+t}) cannot approach to a thermal equilibrium state
because it maintains the initial-state dependence at
$t\!\rightarrow\!\infty$, see Eqs.~(\ref{n+}-\ref{s+}). Correspondingly, the
entanglement dynamics undergoes a quantum phase transition as
the system-environment coupling varying from the weak-coupling
regime to the strong-coupling regime.

\vspace{0.5cm}

\begin{figure}[ht]
\centerline{\scalebox{0.35}{\includegraphics{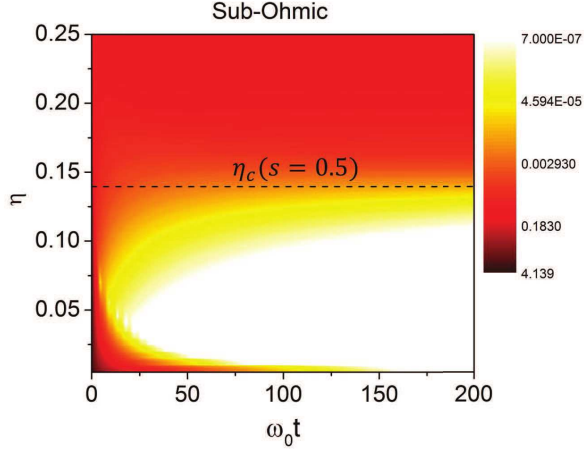}}}
\caption{{\bf Real-time entanglement dynamics}. The contour plot of the entanglement $E_{N}(\rho_{+}(t))$ (in log scale)
 as a function of the time and the system-environment coupling strength $\eta$ at zero initial environment temperature for 
 sub-Ohmic spectral density. The other parameters are taken as $\omega_{c}=3\omega_{0}$, $\kappa=0.5\omega_{0}$ and $r=3$. }
\label{fig1}
\end{figure}

\noindent \textbf{Nonequilibrium quantum phase transition}.
Now, we shall numerically analyze the entanglement dynamics for different spectral densities
to demonstrate the quantum phase transition discussed above. Taking the sub-Ohmic
reservoir $(s=0.5)$ as an example, we present the real-time dynamics of the entanglement in Fig.~\ref{fig1}.
As it shows, in the weak-coupling regime  $\eta<\eta_{c}(s)=0.141$ for $s=0.5$, the entanglement
of the center-of-mass motion  is gradually decohered
away. This is indeed true for all the three different Ohmic-type spectra given by Eq.~(\ref{osp}).
However, in the strong-coupling regime $\eta>\eta_{c}(s)$, the entanglement can be
partially preserved, even in the long-time limit, due to the existence of the localized
state which induces the long-time non-Markovian memory effect.
This provides the real-time dynamics of the nonequilibrium quantum phase transition through entanglement
decoherence as the system-environment coupling strength varies.

\begin{figure}[ht]
\centerline{\scalebox{0.60}{\includegraphics{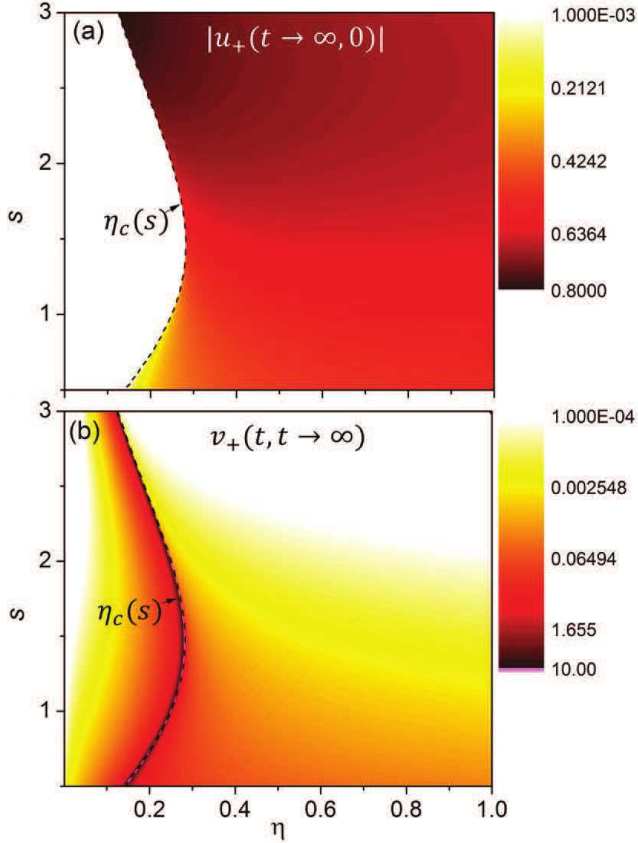}}}
\caption{{\bf Localization and fluctuation dynamics}. (a) The localization given by the retarded 
Green function $|u_{+}(t\rightarrow\infty, 0)|$, and (b) the
fluctuation in terms of the correlated Green function $v_{+}(t,t\rightarrow\infty)$ (in log scale) 
as a function of the coupling strength $\eta$ and the spectral parameter $s$. The other
parameters are taken as the same as in Fig.~\ref{fig1} but the initial environment temperature
$T=0.1\omega_0$. }
\label{fig2}
\end{figure}

To understand the origin of the quantum phase transition, we present the quantum
dissipation and fluctuations, described by the steady-state retarded Green function
$u_{+}(t \rightarrow \infty,0)$ and correlation Green function $v_{+}(t,t \rightarrow \infty)$, in Fig.~\ref{fig2}.
It shows that in the weak-coupling regime ($\eta<\eta_{c}(s)$),  the amplitude of the retarded
Green function $|u_{+}(t\rightarrow \infty, 0)|$ decays to zero, see the empty area in
Fig.~\ref{fig2} (a).  In the strong-coupling regime ($\eta>\eta_{c}(s)$), the localized mode occurs,
then the amplitude of $u_{+}(t\rightarrow \infty, 0)$ maintains a nonzero value,
given by the color area in Fig.~\ref{fig2} (a).  This demonstrates clearly a quantum phase transition
from dissipation dynamics to localization dynamics when the system-environment coupling passes
through the critical coupling $\eta_{c}(s)$
for various spectral densities (different $s$). This phase transition is the first-order
phase transition and is purely induced by quantum fluctuations.
On the other hand, the corresponding steady-state fluctuation Green function $v_{+}(t,t\rightarrow \infty)$
presented in Fig.~\ref{fig2} (b) shows a huge amount of thermal fluctuations near the
quantum critical transition line $\eta_{c}(s)$ (the dash line). When the coupling strength
goes away from the critical regime around $\eta_{c}(s)$, the fluctuations decrease rapidly.
This manifests the quantum criticality as a result of the competition between quantum fluctuations
and thermal fluctuations.

\begin{figure}[h]
\centerline{\scalebox{0.43}{\includegraphics{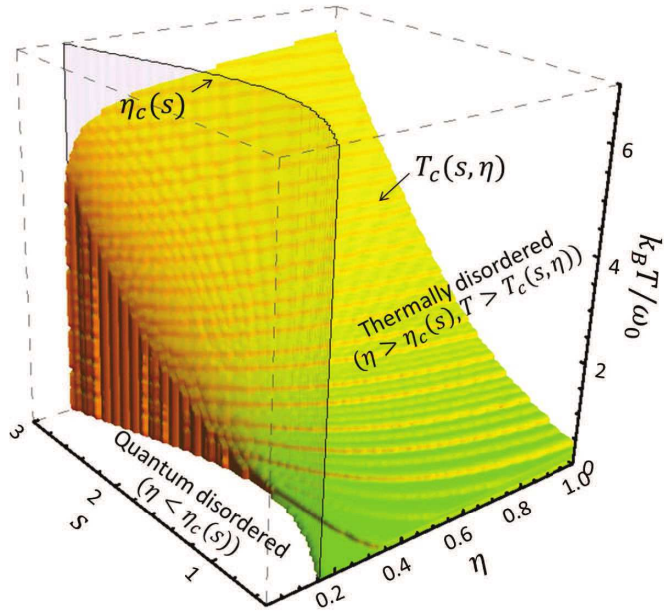}}}
\caption{{\bf The 3-D phase diagram}. A contour plot of entanglement degree
$E_N(\rho_+(t\rightarrow \infty))$ in 3-D space of the coupling strength $\eta$, the spectral parameter $s$
and  the initial reservoir temperature $T$. The other parameters are taken as the same as in Fig.~\ref{fig1}.}
\label{fig3}
\end{figure}

With the above quantum criticality extracted from the fluctuation Green function $v_+(t,t)$, an
entanglement phase diagram, in terms of the entanglement $E_N(\rho_+(t\rightarrow \infty))$
as a function of the system-environment coupling $\eta$,
the spectral parameter $s$ and the initial environment temperature $T$, is presented in Fig.~\ref{fig3}.
As we see, at zero temperature, the $\eta-s$ plane
is separated into the quantum disordered ($\eta <\eta_c(s)$) and quantum ordered ($\eta > \eta_c(s)$) phases.
When the initial environment temperature is nonzero, the competition between quantum fluctuations
and thermal fluctuations shows up.  As a result, the entanglement protected by the localization
in the strong-coupling regime can be gradually decohered away by thermal fluctuations, and the thermal disordered
phase with $E_N(\rho_+(t\rightarrow \infty))=0$ is formed for $\eta > \eta_c(s)$ and $T > T_c(s, \eta)$, where
$T_c(s,\eta)$ is a critical surface separating the quantum ordered phase and
 the thermal disordered phase, as shown in Fig.~\ref{fig3}.
Owing to the strong thermal fluctuation and small localized mode amplitude for the small
value $s$, the thermal disordered phase starts to show up from  the sub-Ohmic reservoir.
Increasing the initial environment temperature will enhance thermal fluctuations such that the
domain of the thermal disordered phase is enlarged, extending to the Ohmic
and then super-Ohmic reservoir.
It also shows that the transition from the quantum entangled phase to the thermal disordered phase is a
continuous phase transition.

\vspace{0.5cm}

\noindent \textbf{\large Discussions}.

In this work, we find that entanglement decoherence, due to the
environment-induced dissipation, localization and fluctuation dynamics, forms
three different types of phases as the system-environment coupling
strength $\eta$, the spectral parameter $s$ and the initial
environment temperature $T$ vary: the dissipation-induced
disentangled phase (phase I) in the weak-coupling regime $\eta
<\eta_c(s)$ for arbitrary initial environment temperature; the
quantum entangled phase (phase II, the colored part in
Fig.~\ref{fig3}) protected by the localized state in the
strong-coupling regime $\eta >\eta_c(s)$ with $T < T_c(s, \eta)$,
and the thermal disordered phase (phase III) as the result of
thermal fluctuations dominating in the regime $\eta >\eta_c(s)$ and $T
>T_c(s, \eta)$. The transition from phase I to phase II is the
first-order quantum phase transition (corresponding to the
transition of dissipation dynamics to localization dynamics
\cite{Lo2015,Nan2015}), while transition from phase II to 
phase III is a continuous phase transition.  The results presented in this article 
can be applied to other open systems \cite{Zhang2012}.  This provides a general
procedure to explore nonequilibrium quantum phase transition through
the experimental measurement of real-time entanglement decoherence dynamics in many-body systems.

\vspace{0.5cm}

\noindent \textbf{Acknowledgments.}

\noindent
We thank Dr.~Ping-Yuan Lo for some useful discussions.
This research is supported by the Ministry of Science and Technology of ROC under
Contract No. NSC-102-2112-M-006-016-MY3. It is also supported in part by the Headquarters
of University Advancement at the National Cheng Kung University,
which is sponsored by the Ministry of Education of ROC.

\ \\
\noindent \textbf{Author Contributions}

\noindent
Y. C. L. and P. Y. Y. performed the theoretical calculations and prepared all the figures;
W. M. Z. proposed the ideas and wrote the main manuscript. All authors participated
the discussions and revised the manuscript.

\ \\
\noindent\textbf{Competing financial interests}

\noindent The authors declare no competing
financial interests.

\ \\
\noindent\textbf{Additional Information}

\noindent Correspondence and requests
for materials should be addressed to W. M. Zhang \\ (wzhang@mail.ncku.edu.tw).

\end{document}